# Covid-19 classification with deep neural network and belief functions

Ling Huang[1*], Su Ruan [2*], and Thierry Denoeux[1,3]

[1]Université de technologie de Compiègne, CNRS, Heudiasyc, Compiègne, France
[2]Université de Rouen, Quantif, LITIS, Rouen, France
[3]Institut universitaire de France, Paris, France

**Abstract.** Computed tomography (CT) image provides useful information for radiologists to diagnose Covid-19. However, visual analysis of CT scans is time-consuming. Thus, it is necessary to develop algorithms for automatic Covid-19 detection from CT images. In this paper, we propose a belief function-based convolutional neural network with semi-supervised training to detect Covid-19 cases. Our method first extracts deep features, maps them into belief degree maps and makes the final classification decision. Our results are more reliable and explainable than those of traditional deep learning-based classification models. Experimental results show that our approach is able to achieve a good performance with an accuracy of 0.81, an F1 of 0.812 and an AUC of 0.875.

## 1 Introduction

The Covid-19 pandemic continues to have a big influence on the health and daily life of the global population. The most important step for stopping Covid-19 is to detect infected patients effectively and impose immediate isolation. Patients infected by Covid-19 were found to present abnormalities in CT images, which makes it possible to detect Covid-19 cases in clinical medicine. Also, the degree of infection can be assessed from CT images, which is helpful for proper treatment and care. Recently, a lot of deep learning-based Covid-19 case detection methods have been proposed, some of which have been quite successful. However, the limited quantity of training cases and annotations is still the main challenge for improving the classification accuracy of Covid-19. Furthermore, due to the low contrast of CT images, deep learning-based classification methods have difficulty in processing uncertain and imprecise information, such as the pixels close to boundary and images pertaining to non-Covid-19 cases. To solve the above problems and thanks to the open source Covid-19 dataset from the research community, we propose a belief function-based classification network to classify Covid-19 cases using semi-supervised learning. The contribution of this paper can be summarized as follows: 1) we propose a novel neural network architecture for Covid-19 detection from CT images composed of a convolutional network part for feature extraction and a belief function-based classification module; 2) the evidential neural network classifier maps image features into mass functions for improved

---

* Corresponding authors: ling.huang@hds.utc.fr, su.ruan@univ-rouen.fr

Covid-19 case classification; 3) a semi-supervised training algorithm is proposed to train the framework with partial annotations. We show that this work can contribute to more reliable detection of Covid-19 cases and make the best use of relevant information from CT images.

## 2 Related Work

### 2.1 Covid-19 Datasets

Although a lot of research has been done on the Covid-19 pandemic, only a few datasets are publicly available, due to privacy protection concerns and the delay in collecting information. The Italian Society of Medical and Interventional Radiology (SIRM) provided chest X-rays and CT images of 68 Italian Covid-19 cases [1]. Murphy et al. released a dataset of chest X-rays and CT images from 99 Covid-19 cases at Radiopaedia [2]. Cohen et al [3] released a Covid-19 dataset that contains 45 patients with 84 Covid-19 X-rays images. Yang et al [4] built a COVID-CT dataset with 349 Covid-19 CT images from 216 patients and 397 non-Covid-19 CTs. These data are very useful for the development of automatic detection methods.

### 2.2 Deep Learning-Based Diagnosis of COVID-19

Since the Covid-19 outbreak, a lot of efforts have been devoted to the development of automatic deep learning methods to perform screening of Covid-19 cases from medical images [6-10]. For the analysis of CT scans, Yang et al. [4] proposed a DesNet-based Covid-19 case classification model; He et al. introduced an EfficientNet [5]-based deep model for Covid-19 diagnosis [6]; Xu et al. developed a multiple CNN-based screening model to classify patients with Covid-19 [7]. Alternatively, Wang et al. [8] proposed a tailored deep convolutional neural network (COVID-Net) for detection of Covid-19 cases from chest X-ray images using EfficientNet and an additional lung dataset. Also, some researchers carried out deeper analysis of Covid-19 data. Zhou et al. [9] described an automatic Covid-19 CT segmentation framework using a U-Net integrating spatial and channel attention mechanisms. Amyar et al. [10] proposed a multi-task model for both Covid-19 CT classification and segmentation by putting together an encoder, a decoder and a classifier. All of these existing deep learning-based Covid-19 screen methods are mainly based on the probabilistic formalism, which is more limited than the theory of belief function introduced in the next section.

### 2.3 Belief Function Theory

#### 2.3.1 *Representation of Evidence*

The theory of belief functions, also known as *Dempster-Shafer theory* or *evidence theory*, was first introduced by Dempster [11] and Shafer [12] and further popularized and developed by Smets [13]. This theory is a generation of the Bayesian theory, but it is more flexible and it allows for a wider range of uncertain and imprecise information. The great expressiveness belief function theory makes it possible to represent evidence in a more faithful way than using probability theory. Let $\Omega = \{\omega_1, \omega_2, \dots, \omega_K\}$ be a finite set of hypotheses about some question. Evidence about $\Omega$ can be represented by a mapping $m$ from $2^\Omega$ to [0,1] such that $\sum_{A \subseteq \Omega} m(A) = 1$, called a *mass function*. For any hypothesis $A \subseteq \Omega$, the quantity $m(A)$ represents the mass of belief specifically allocated to $A$, and that cannot be allocated to any strict subset. Given a mass function $m$, belief and plausibility functions from $2^\Omega$ to [0,1] can be defined, respectively, as

$$Bel(A) = \sum_{B \subseteq A} m(B), \quad (1)$$
$$Pl(A) = \sum_{B \cap A \neq \emptyset} m(B) = 1 - Bel(\bar{A}), \quad (2)$$

for all $A \subseteq \Omega$. Quantity $Bel(A)$ can be interpreted as the degree of evidence that totally support $A$, while $Pl(A)$ can be interpreted as the degree of evidence that does not contradict $A$.

### 2.3.2 Dempster's Rule for Evidence Fusion

Two mass functions $m_1$ and $m_2$ derived from two independent items of evidence can be combined by *Dempster's rule* [11] defined as

$$(m_1 \oplus m_2)(A) = \frac{1}{1-\kappa} \sum_{B \cap C = A} m_1(B) m_2(C), \quad (3)$$

for all $A \subseteq \Omega, A \neq \emptyset$, and $(m_1 \oplus m_2)(\emptyset) = 0$. In Eq. (3), κ represents the *degree of conflict* between $m_1$ and $m_2$, defined as

$$\kappa = \sum_{B \cap C = \emptyset} m_1(B) m_2(C). \quad (4)$$

## 3 Methods

### 3.1 Overview of the Architecture

Inspired by the Evidential Neural Network (ENN) model introduced in [14], we propose here an evidential Covid-Net for Covid-19 case classification. Figure 1 shows the proposed architecture. The input of this framework is a Covid-19 image. Firstly, ResNet50 [15] is used as for feature extraction; we take the extracted features from the last convolution layer. Secondly, an adaptive average pooling layer is used after feature extraction to get a 2048-dimensional feature vector. Then a linear layer is added after the pooling layer in order to reduce the number of features from 2048 to 64. The selection of 64 is from experience. We took efficiency and computation cost into consideration and reduced the number of features from 2048 into 64. Lastly, an evidential module is added after the linear layer to map image features into mass functions used for classification. The structure of the evidential module is similar to that of the ENN model, comprised of three parts: a distance computation layer, a basic belief assignment layer and a mass fusion layer in which mass functions are combined by Dempster's rule. Details of the evidential module will be introduced in Section 3.2. Finally, the network output is a decision to classify the input image as a Covid-19 case or not.

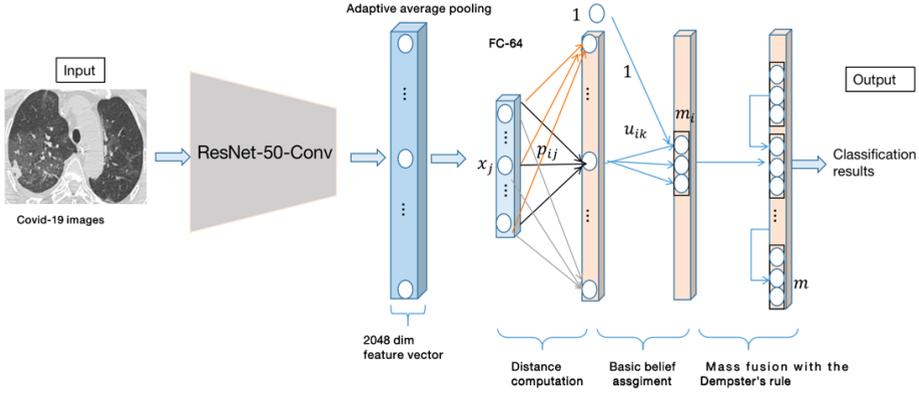

**Fig. 1.** Evidential Covid-Net architecture.

### 3.2 Evidential Neural Network Classifier

Neural networks have achieved remarkable performances in machine learning tasks and have become very popular, but their black-box nature makes them ill-suited to decision-aid in the medical domain. In [14], Denoeux first proposed an Evidential Neural Network (ENN) classifier, which outputs mass functions for classification tasks. The ENN classifier first summarizes the input features as $r$ prototypes initialized, e.g., by the K-means algorithm. Each prototype $p_i$ is a piece of evidence about the class of input $x$, whose reliability decreases with the distance $d_i$ between $x$ and $p_i$. Mass function induced by prototype $p_i$ is

$$m_i(\{\omega_k\}) = \alpha_i u_{ik} exp(-\gamma_i d_i^2), \qquad (5)$$
$$m_i(\Omega) = \alpha_i exp(-\gamma_i d_i^2), \qquad (6)$$

where $u_{ik}$ is the degree of membership of prototype $p_i$ to class $\omega_k$, i = 1, ... r, k = 1,2, and $\alpha_i$ and $\gamma_i$ are two tuning parameters. The evidences from the $r$ prototypes are then combined by the Dempster's rule

$$m = \oplus_{i=1}^{r} m_i. \qquad (7)$$

The parameters to be learned during training are the prototypes $p_i$, the membership degrees $u_{ik}$, $\alpha_i$ and $\gamma_i$. The learning process can be conducted by minimizing the cost function

$$C(\psi) = \sum_{i=1}^{r}\sum_{k=1}^{c}(pl_{ik} - y_{ik})^2 + \lambda \sum_{i=1}^{r} \alpha_i, \qquad (8)$$

where $\psi$ is the vector of all parameters, $pl_{ik}$ is the output plausibility and the ground truth for instance $i$ and class $k$, and $\lambda$ is a regularization coefficient. The regularization coefficient $\lambda$ can be either determined by cross-validation or fixed to a given value chosen from experience. Here we set $\lambda$=0.01 from experience as suggested in [14]. The details of the optimization algorithm are described in [14].

### 3.3 Loss functions for Semi-Supervised Learning

The need for big annotated training data has become a bottleneck in deep learning, which limits its application in the medical domain where the labelling task requires not only careful delineation but also high-level professional knowledge. Thus, in this paper we propose a semi-supervised learning method for Covid-19 case classification. For each input image $x$, we use several transformations to get new images, noted $x_t$. Similar images are expected to produce similar classification results even if some transformations have been performed on them.

Two loss functions are proposed for training images with label and without label. Here we train the network with 50% images and their corresponding labels using the following loss1 function,

$$loss1 = -y\log(m(\{\omega_1\})) - (1-y)\log(m(\{\omega_2\})), \quad (9)$$

which measures the difference between the output mass $m$ and the ground truth label y. For the remaining 50% images, we do not use label information. We calculate the mass functions for the image $x$ and the transformed image $x_t$ and we minimize their distance using the following loss2 criterion, which measures the difference between the origin output and the transformed output:

$$loss2 = \sum_{t=1}^{T}\left(m(\{\omega_1\}) - m_t(\{\omega_1\})\right)^2 + \left(m(\{\omega_2\}) - m_t(\{\omega_2\})\right)^2, \quad (10)$$

where m is the mass function of the input image $x$, y is the classification label, $m_t$ is the mass function of the transformed image $x_t$, and T is the number of transformed images.

## 4 Experiment

### 4.1 Dataset

In this paper we choose the COVID-CT dataset [4] to test our proposal because it contains diverse information from 216 patients. Out of the 746 instances, 425 were used for training (including 191 positive cases), 118 were used for validation (including 60 positive cases), and 203 were used for testing (including 98 positive cases). Table 1 and Table 2 show the composition of COVID-CT dataset. For the training images, we applied the following preprocessing operations: resizing to 256*256, random resizing and cropping to 224*224 and normalization. Similarly, validation and test images were preprocessed as follows: resizing to 256*256; center cropping to 224*224 and normalization. We also used some image transformations to generate new images for semi-supervised training. Based on the preprocessed image $x$, we added Gaussian noise and flipped images with horizontal parameter 0.5 to get a new image $x_t$. During training, an early-stop mechanism is used during training. If the performance does not increase in 5 iterations, the training will be stopped. We used three metrics to evaluate our proposal: accuracy, F1, and area under ROC curve (AUC). For all three metrics, the higher the better. **Fig. 2** shows some examples of images from the COVID-CT dataset.

Table 1. The composition of COVID-CT dataset (number of images)

| Type | Non-Covid-19 | Covid-19 | Total |
| --- | --- | --- | --- |
| train | 234 | 191 | 425 |
| validation | 58 | 60 | 118 |
| test | 105 | 98 | 203 |

Table 2. The composition of COVID-CT dataset (number of patients)

| Type | Non-Covid-19 | Covid-19 | Total |
| --- | --- | --- | --- |
| train | 105 | 1-130 | 235 |
| validation | 24 | 131-162 | 56 |
| test | 42 | 163-216 | 96 |

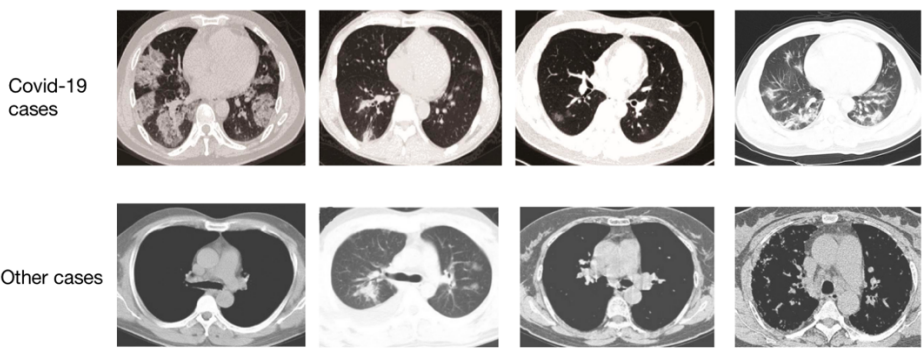

**Fig. 2.** Examples of images from the COVID-CT dataset. The first row contains images with Covid-19 cases and the second row contains images with other cases.

### 4.2 Results

We compared our results with those of the baseline model ResNet and other state-of-the-art methods on the Covid-19 dataset. **Table 3** shows the results after training the models with labeled images. We first compared our proposal with the baseline method ResNet50 and obtained 2.5%, 2.6% and 3.5% increase in, respectively, Accuracy, F1 and AUC. Compared with the DensNet-based model Covid-CT-Net, we get 1%, 0.3% and 0.6% increase in Accuracy, F1 and AUC, respectively. Our performance does not reach the performance of the best method Covid-Net, but the results are not strictly comparable because Covid-Net used additional information in the lung dataset for self-training. As for semi-supervised training, **Table 4** shows the results of models trained with all learning images but only 50% of the labels. Compared with the baseline method ResNet50, our proposal obtained, respectively, 6.8%, 12.1%, 6.4% increase in Accuracy, F1 and AUC. Compared with Covid-CT-Net, our proposal obtained, respectively, 3.3%, 4.1%, 1% increase in Accuracy, F1 and AUC. For semi-supervised training, our proposal shows better performance than the baseline ResNet50 method and the state-of-the-art method Covid-Net. Compared with fully supervised training (using all training labels), semi-supervised training (with 50% of labels used) achieved comparable performance, with only 7.9%, 11%, 0.5% decrease in, respectively, Accuracy, F1 and AUC. Fig. 3 shows the ROC curves of our method with fully and partially supervised training.

**Table 3.** Classification results for the COVID-CT test dataset (100% training labels)

| Methods | Accuracy | F1 | AUC |
| --- | --- | --- | --- |
| ResNet50 | 0.785 | 0.786 | 0.84 |
| Covid-CT-Net [4] | 0.8 | 0.809 | 0.869 |
| Covid-Net [6] | **0.85** | **0.86** | **0.94** |
| Evidential Covid-Net | 0.81 | 0.812 | 0.875 |

**Table 4.** Classification results for the COVID-CT test dataset (50% training labels)

| Methods | Accuracy | F1 | AUC |
| --- | --- | --- | --- |
| ResNet50 | 0.663 | 0.581 | 0.806 |

| Covid-CT-Net [4] | 0.698 | 0.661 | 0.869 |
| Evidential Covid-Net | **0.731** | **0.702** | **0.870** |

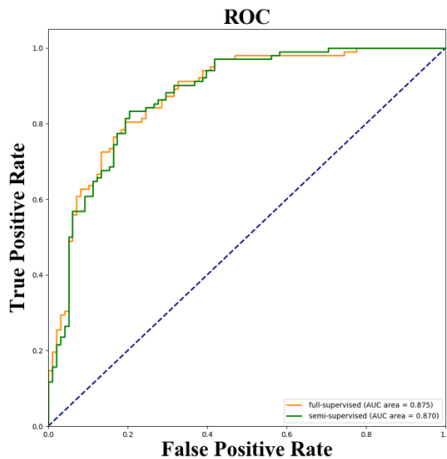

Fig. 3. ROC curves with full-supervised training (orange) and semi-supervised training (50% training labels, green).

## 5 Conclusion

We have proposed an evidential Covid-Net model for automatic Covid-19 detection. Our model combines the ResNet architecture with an evidential layer to map high-level image features into mass functions. The belief function formalism offers a new approach for reasoning with pieces of evidence and make the results more easily interpretable than those of probabilistic methods. Moreover, we proposed a semi-supervised training method to train the evidential Covid-Net model. In future research, we will investigate the fusion of classification results from different image modalities as well as additional information in the belief function framework, to enhance the detection performance. Also, we will extend our approach to more powerful CNN architectures such as EfficientNet to obtain higher classification accuracy.